\documentclass[12pt]{article}                                                %
\usepackage{amsmath,amssymb,amsfonts,amsbsy}                                 %
\usepackage{cite}                                                            %
\usepackage{graphicx}                                                        %
\usepackage{wrapfig}                                                         %
\usepackage{epsfig}                                                          %
\usepackage{bbm} 
\usepackage{bm} 
\usepackage{color}                                                       
\usepackage{dsfont} 
\usepackage{latexsym} 
\usepackage{lscape} 
\usepackage{mathrsfs} 
\usepackage{morefloats} 
\usepackage{floatflt} 
\usepackage{slashed} 
\usepackage{psfrag}                                                          %
\DeclareFontFamily{OT1}{mygreek}{}
\DeclareFontShape{OT1}{mygreek}{m}{n}{<->omsegr}{}
\DeclareFontShape{OT1}{mygreek}{b}{n}{<->omsegrb}{}
\DeclareFontShape{OT1}{mygreek}{m}{it}{<->omsegri}{}
\DeclareFontShape{OT1}{mygreek}{bx}{n}{<->sub * mygreek/b/n}{}
\DeclareFontShape{OT1}{mygreek}{m}{sl}{<->sub * mygreek/m/it}{}
\DeclareSymbolFont{Greekrm}{OT1}{mygreek}{m}{n}                              %
\DeclareSymbolFont{Greekbf}{OT1}{mygreek}{b}{n}                              %
\DeclareSymbolFont{Greekit}{OT1}{mygreek}{m}{it}                             %
\DeclareMathSymbol{\omegab}{\mathalpha}{Greekbf}{119}                        %
\begin{document}                                                             %

\setcounter{section}{0} \setcounter{subsection}{0} 
\setcounter{equation}{0} \setcounter{figure}{0} 
\setcounter{footnote}{0} \setcounter{table}{0}

\begin{center}
\textbf{SUMMARY. AN ANALYTICAL REVIEW OF DSPIN-13} \vspace{5mm}

A.V. Efremov$^{1,\dag}$ and 
{J. Soffer}$^{2,\ddag}$ 

\vspace{5mm}

\begin{small}
  $^1$\emph{JINR, Dubna, Russia} \\
  $^2$\emph{Temple University, Philadelphia, USA} \\
  $\dag$ \emph{E-mail: efremov@theor.jinr.ru}
  $\ddag$ \emph{E-mail: jacques.soffer@gmail.com}\\
\end{small}
\end{center}

\vspace{10.0mm} 

The XV Workshop on High Energy Spin Physics continued a series of 
meetings, the first of which was held in Dubna in 1981 on the 
initiative of a prominent theoretical physicist L.I.~Lapidus. 
Since then, such meetings are taking place in Dubna each odd 
year. They give the opportunity to present and discuss the 
accumulated annual news. Another important feature is the 
possibility of participation for a large number of physicists 
from the former Soviet Union and Eastern Europe, for whom 
long-distance travel is difficult by financial (and in past as 
well by bureaucratic) reasons. In even years large International 
Symposia on Spin Physics have been held in various countries, 
including Dubna, Russia in 2012.
         
This meeting was characterized by a substantial attendance, with 
a larger than ever number of participants (125 persons) from 
different countries: Russia 24, USA 10, Belarus 7, Poland 6,  
Germany 4, Czech Republic 3, Italy 3, France 2, Slovakia 2, Iran 
2, China 2 and by one person from Belgium, Bulgaria, India, 
Portugal, Sweden, Ukraine and South Korea. As always, a lot of 
physicists from JINR (53) were involved.

The reason for the increasing popularity of the meeting is, 
apparently, the fact that this year has brought many new 
experimental results and above all the discovery and 
determination of the quantum numbers of the Higgs boson at the 
Large Hadron Collider (LHC), given in talks by A. Rinkevicius 
(USA) and Yaquan Fang (China).
    
The talk by X.~Artru (France) proposed the development of simple 
explanation of the Collins effect and the effect of handedness in 
the model of sequential fragmentation of quark and offered a 
program of implementation of the model into Monte Carlo 
simulation.

Classical experiments on the study of the nucleon spin structure 
at high energies use both scattering leptons on polarized 
nucleons (HERMES, JLab, COMPASS) and collisions of the polarized 
protons (RHIC, IHEP, JINR). The joint description of such 
different high-energy processes becomes possible due to the 
application of the fundamental theory of strong interactions, 
quantum chromodynamics (QCD), and its remarkable properties of 
factorization, local quark-hadron duality and asymptotic freedom 
which allow one to calculate the characteristics of a process 
within the framework of perturbation theory (PT). At the same 
time, parton distribution functions (PDF), correlations and 
fragmentation functions are universal and do not depend on the 
process. However since they are not given by the theory, they 
require some methods to build specific models. A number of 
reports at the conference were dedicated to the development and 
application of this type of models (P.~Zavada, Czech Republic, 
the original covariant model of nucleon, J.~Soffer, France, 
quantum statistical model and others). Several talks were devoted 
to the development of methods of experimental data processing and 
extraction both polarized and unpolarized PDF. It is particularly 
worthwhile mention the report of D.~Str\'{o}zik-Kotlorz 
(Poland) on the development of the method of truncated Mellin 
moments and generalized evolution equations for these moments, 
and the talk of A.~Sidorov (Dubna) who demonstrated the 
particular importance of the knowledge of quark fragmentation 
functions for the determination of spin dependent PDFs of sea 
quarks. New data of COMPASS collaboration on measurement of quark 
fragmentation functions into pions and kaons were presented by N. 
du Fresne von Hohenesche from Mainz.

The theoretical description of processes involving spin, 
especially dependent on internal transverse parton motion (TMD), 
proves to be, as always, more complicated, so that the number of 
these functions increases and the picture connected with them 
loses to a considerable degree the simplicity of a parton model 
with its probabilistic interpretation. One of the difficulties 
here is the evolution of these functions with a change in the 
wavelength of a tester. A possible approach to its solution was 
presented in the talk by I.~Cherednikov (Dubna and Antwerpen).

The most widely studied to date is the helicity distribution of 
quarks in the nucleon $g_{1}$. The COMPASS data (A.~Ivanov, 
Dubna) allow one essentially specify these distributions. New 
measurements of the structure function $g_{2}$ of the proton and 
neutron ($^{3}$He) were presented by Jian-ping Chen from JLab. 
They show better agreement with the so-called Wandzura-Wilczek 
approximation relating these distributions at leading-twist. 
Recent experimental data are precise enough to include in their 
QCD analysis not only the perturbation corrections, but also the 
contributions of higher-twist and target mass corrections 
(F.~Arbabifar and F.~Abdolghafari, Tehran). In this case, 
positive polarization of strange quarks is excluded with high 
probability. New data on the spin distributions of sea $\bar u$ 
and $\bar d$ quarks from the $W^{+}$ and $W^{-}$  bosons 
production processes in polarized proton-proton collision were 
presented by the STAR collaboration (K.~Barish, BNL), in good 
agreement with the predictions of the statistical model 
(J.~Soffer). The polarization of gluons, however, is consistent 
with the results of its direct measurements by the COMPASS and 
PHENIX + STAR collaborations (K.~Barish -- BNL,  Qinghua Xu -- 
China). Its low value seems insufficient for resolving the 
so-called nucleon spin crisis.

The hope to overcome the crisis lies with contributions of the 
orbital angular momenta of quarks and gluons which can be 
determined by measuring the so-called Generalized Parton 
Distributions (GPD). Theoretical aspects of a model GPD 
calculation were covered in talks of S.~Goloskokov (Dubna) and 
S.~Nair (Bombay). Different experimental aspects of GPD 
measurement already held (HERMES) and new ones under preparation 
(COMPASS) were presented by W.-D.~Nowak (Freiburg) and A.~Sandacz 
(Warsaw), respectively.

Other important spin distribution functions manifest themselves 
in scattering of transversely polarized particles. The processes 
in which the polarization of only one particle (initial or final) 
is known are especially interesting and complicated from the 
theoretical point of view (and relatively simple from the point 
of view of experiment -- such complementarities frequently 
occur). Such single spin asymmetries are related to the T-odd 
effects, i.e. they seemingly break invariance under time 
reversal. Here, however, we deal with an effective breaking 
connected not with the true non-invariance of fundamental (in our 
case, strong, described by QCD) interactions under the time 
reversal, but with their simulations by thin effects of 
rescattering in the final or initial state. 

The effects of single spin asymmetries have been studied by 
theorists (including Dubna theorists who have priority in a 
number of directions) for more than 20 years, but their study 
received a new impetus in recent years in connection with new 
experimental data on the single spin asymmetry in the 
semi-inclusive electro-production of hadrons off longitudinally 
and transversely polarized targets at the facilities COMPASS 
(F.~Bradamante, A.~Bressan, Trieste), HERMES (W.-D.~Nowak, 
Freiburg) and CLAS (Jian-ping Chen, Newport News). In particular, 
data from HERMES for the asymmetry of pions (the so-called 
"Sivers function") associated with the left-right difference in 
the distribution of partons in a transversely polarized hadron 
are described by the existing theory. However, the data for 
positive kaons in the region of small $x$ about 2.5 times deviate 
from their predictions. New measurements of the asymmetry by the 
COMPASS collaboration give evidence in favor of the explanation 
of this difference by higher-twist contributions. Especially 
interesting was the comparison of SSA (transversity) in the 
production of a pair ($\pi^{+}$,$\pi^{-}$) from transversally 
polarized proton: $x$-dependence of the pair is almost identical 
to the $x$-dependence of  $\pi^{+}$ (F.~Bradamante, Trieste), 
which clearly testifies to a sequential fragmentation mechanism 
proposed by X.~Artru. New data on the SSA pions produced in 
polarized proton-proton collisions at RHIC energies 
(200$\times$200 GeV) were provided by the STAR collaboration 
(Qinghua Xu, Shandong). The collaboration confirms surprisingly 
large asymmetries observed previously at lower energies, which 
indicates their energy independence. However, new measurements at 
large $p_{T}$ show also that asymmetry has roughly constant 
behavior up to $p_{T}=7$ GeV/c. This creates great difficulties 
for the modern theoretical understanding of these processes. New 
data were also obtained for the asymmetry of pairs of hadrons 
($\pi^{+}\pi^-$), which creates opportunities for measuring the 
PDF transversity (distribution of transversely polarized quarks 
in a transversely polarized nucleon). Similar observations were 
reported by the PHENIX collaboration (K.~Barish, BNL). Also, 
PHENIX does not see much difference in the asymmetries of 
$\eta^{-}$ and $\pi^{0}$ mesons earlier reported by STAR. The 
specific mechanisms of origin of these asymmetries still remain a 
mystery. Thus, although in general the single asymmetry is 
described by the existing theory, its development continues. 
Appearing here T-odd PDF lose key properties of universality and 
become "effective" depending on the processes in which they are 
observed. In particular, the most fundamental prediction of QCD 
is a change in the sign of the Sivers function in the transition 
from pion electroproduction to the production of Drell-Yan pairs 
on a transversely polarized target. This conclusion is supposed 
to be checked in the COMPASS experiment (O.~Denisov, Turin) and 
at colliders RHIC, NICA (R.~Akhudzyanov, Dubna) and PANDA-PAX. We 
also had a very interesting talk on the importance of the 
Drell-Yan process and an ongoing experiment to improve our 
knowledge of the flavor structure of the nucleon sea (Jen Chieh 
Peng, Illinois).

Considerable interest and discussion were called by new data of 
the JLab on measurement of the ratio of the electric and magnetic 
form factors of the proton carried out by "technique of the 
recoil polarization" presented at the meeting (Ch.~Perdrisat -- 
Williamsburg, V.~Punjabi -- Norfolk State University). Early 
measurements of the JLab showed that this ratio is not constant, 
as it has been believed for a long time, and decreases linearly 
with increasing momentum transfer $Q^2$ (the so-called "form 
factor crisis"). New data obtained in 2010 (experiment GEp(3) 
with JINR participation), point to a flattening of this ratio in 
$Q^{2}=6-8$ GeV$^{2}$. The proposed experiment GEp(5) will 
advance up to $Q^{2 }=15-17$ GeV$^{2}$. The question whether this 
is the behavior due to the influence of radiative corrections, in 
particular, two-photon exchange, is still open.

Several talks were devoted to theoretical search of $Z'$ features 
and other exotic at the LHC and the future International Linear 
Collider (ILC) of electrons and positrons (V.~Andreev, 
A.~Tsitrinov -- Gomel). 

A separate section was devoted to a problem of localization of 
energy momentum and spin in the classical field theory. The 
picture arising in geometrical description in the language of 
external forms is close to the traditional: if electrons (and 
quarks) are described by an initial tensor of energy-momentum, 
for photons (and for gluons!) it is a tensor of Belinfante 
(F.~Hehl, Cologne). This conclusion is very actual in the light 
of the discussed problem of gluon contribution to the nucleon 
spin being discussed. The dynamics of spin in gravitational 
fields and non-inertial reference systems were considered in 
talks by Yu.~Obukhov (IBRAE, Moscow) and A.~Silenko (JINR, 
Dubna). It was shown that particles with spin were the only 
tester for so-called "torsion" of space-time, and their unitary 
transformation allow one to pass to a quasi-classical limit and 
to compare evolutions of quantum spin and classical top. 

Calculation of spin and orbital moment contributions on a lattice 
was discussed in M.~Deka's (JINR, Dubna) talk. In particular, 
essential cancellation of spin and orbital moments of $d$-quarks 
was confirmed.

Finally, considerable attention was paid to the history of 
polarized studies and to further development of the projects of 
polarization studies at FERMILAB (A.~Krisch, Ann Arbor). Plans 
for further research at the modified accelerator Jlab, as well as 
plans to create the electron-nuclear colliders in the world: 
eRHIC, LHeC, MEIC/EIC and especially EIC@HIAF in China were 
presented by Jian-ping Chen, Newport News.

The program of obtaining of polarized proton and antiproton beams 
from the decay of Lambda particles at the U-70 IHEP, Protvino, 
for spin studies at the facility SPASCHARM was presented by 
S.~Nurushev. He stressed the importance of a comparative study of 
spin effects induced by particles and antiparticles.

Of particular interest were plans to create in IKP (J\"{u}lich, 
Germany), a unique European complex for measurement of the 
electric dipole moment (EDM) proton and nuclei (N.~Nikolaev, 
Landau ITP). The fact is that the dipole moment of the 
fundamental particles, if it exists, violates the laws of 
conservation of spatial and temporal parity. Detection of EDM 
would indicate  violation of the Standard Model and, in 
particular, would open up the possibility for an approach to the 
problem of understanding the baryon asymmetry of the Universe. 
The planned complex will lower the measurement limit of deuteron 
EDM up to $10^{-29}$ e$\cdot$cm.

The talks related to the development of the VBLHEP accelerating 
complex of JINR were also presented in the program of the 
conference (V. Ladygin, R. Kurilkin, S. Piyadin, E. Strokovsky -- 
Dubna). They discussed some of the new proposals for research on 
the basis of the upgraded Nuclotron-M. In particular, the 
proposal for a new experiment BM@N whose main purpose is to study 
the properties of dense nuclear matter especially with strange 
quark participation.

Special plenary and parallel sessions were devoted to the project 
of the collider complex NICA at JINR. The project has two phases. 
The first one is the construction of the collider and 
Multi-Purpose Detector (MPD) for studies of heavy ion collisions 
to be completed in 2017. The second phase includes the 
construction of the infrastructure for the acceleration of 
polarized protons and deuterons in the total  energy range 12-27 
GeV with luminosity $\ge $ 10$^{32}$ cm$^{-2}$s$^{-1}$ for 
protons (talk of A.~Kovalenko, Dubna) and a detector for the 
collision products (SPD) reported  by G.~Mescheryakov, Dubna. The 
proposed scheme allows the complex to operate with polarized 
(longitudinal and transversal) or unpolarized proton and deuteron 
beams. The main ideas proposed for the SPD centered around the 
nucleon spin structure using the Drell-Yan process of lepton 
pairs (R.~Akhunzyanov, Dubna), direct photon (A.~Gus'kov, Dubna ) 
and the $J/\Psi$-mesons production. The possibility of 
4$\pi$-geometry of SPD for registration of pairs $e^{\pm}$, 
$\mu^{\pm}$ and direct photons can allow one to measure all 
leading TMD distribution functions of quarks and antiquarks in 
the nucleon. Some of them were measured recently in SIDIS 
experiments, some are still unmeasured. One of the main purposes 
is to check the fundamental QCD predictions for the change of the 
sign of the T-odd TMD in the Drell-Yan process compared with that 
of SIDIS. There were also proposals for the study of spin 
processes in elastic $pp$-scattering (S.~Shimanski and V.~Sharov, 
Dubna), in particular, the so-called "Krisch-effect". Sources of 
polarized particles and physics of acceleration of polarized 
beams (Yu.~Filatov, Dubna, Yu. and M.~Kondratenko, Novosibirsk) 
were discussed. The spin community presented at the meeting 
supported plans for a new and unique opportunities for 
polarization studies at the collider JINR complex. The complex 
with these features will not have competition with other centers 
of polarization studies and the data collected will help to solve 
the riddle of spin effects that has not had solutions since the 
70s of the last century.

Special session on the development of the so-called analytic 
perturbation theory (APT) by Solovtsov-Shirkov was devoted to the 
blessed memory of Alexander P.~Bakulev. As it is known, the 
effective coupling constant in QCD, $\alpha _{s}(Q^{2})$, has a 
non-physical pole in the area of 200-300 MeV (the so-called 
"Landau-Pomeranchuk pole"), which prevents the application of  
QCD perturbation theory in the region of small momentum 
transfers. Imposition of an additional condition on the 
analyticity of divergent series defining $\alpha _{s}(Q^{2})$ 
eliminates the pole and makes the value of $\alpha _{s}(Q^{2})$ 
finite up to $Q^2 = 0$. This leads to noticeable stabilization of 
perturbation theory and to better agreement with experiment up to 
the small $Q^2$, e.g. up to GeV$^{2}$ for the value of $\Gamma^{p 
- n}$ (talks of V.~Khandramai, Gomel). Various aspects of the 
application of this theory as well as a difficult situation in 
QCD description of transition form factor $F_{\gamma \gamma*\pi 
}$ were the subject of talks by O.~Solovtsova  (Gomel),  
A.~Oganesian (ITEP, Moscow ), N.~Stefanis (Bochum), S.~Mikhailov, 
O.~Teryaev, A.~Pimikov and D.~Shirkov (JINR, Dubna) who have had 
a long collaboration with A.P. Bakulev.

The summary of the meeting was made in the final report by 
J.~Soffer. 

The success of the conference was due to the support by the 
Russian Foundation for Basic Research, International Committee 
for Spin Physics, Foundation "Dynasty", European Physical Society 
and the JINR programs for international collaboration: 
Heisenberg-Landau, Bogolyubov-Infeld and Blokhintsev-Votruba 
ones. This made possible to provide noticeable financial support 
to participants from Russia and other JINR Member States and 
developing countries. The materials of the conference, including 
all presented talks
and pdf-file of Proceedings 
are available on the site 
{\tt http://theor.jinr.ru/$\sim$spin/2013/}. 

\end{document}